\PassOptionsToPackage{x11names,table,dvipsnames}{xcolor}
\documentclass[acmtog,timestamp,nonacm]{acmart}

\usepackage{booktabs} 
\usepackage{scalerel}
\usepackage{dsfont}
\usepackage[ruled]{algorithm2e} 
\usepackage{color}
\usepackage{tabularx}

\SetAlFnt{\small}
\SetAlCapFnt{\small}
\SetAlCapNameFnt{\small}
\SetAlCapHSkip{0pt}

\acmJournal{TOG}

\definecolor{redcolor}{rgb}{0.8,0,0}
\definecolor{bluecolor}{rgb}{0,0,0.8}
\definecolor{greencolor}{rgb}{0.0,0.5,0.0}

\newcommand{\Real}{\ensuremath{\mathbb{R}}}
\newcommand{\V}{\ensuremath{V}} 
\newcommand{\T}{\ensuremath{T}} 
\newcommand{\F}{\ensuremath{F}} 
\newcommand{\xun}{\ensuremath{\bar{\mathbf{x}}}} 
\newcommand{\xd}{\ensuremath{\mathbf{x}}} 

\hypersetup{draft}
\begin{document}
\title{Diffusion Structures for Architectural Stripe Pattern Generation}

\begin{teaserfigure}
  \includegraphics[width=\textwidth]{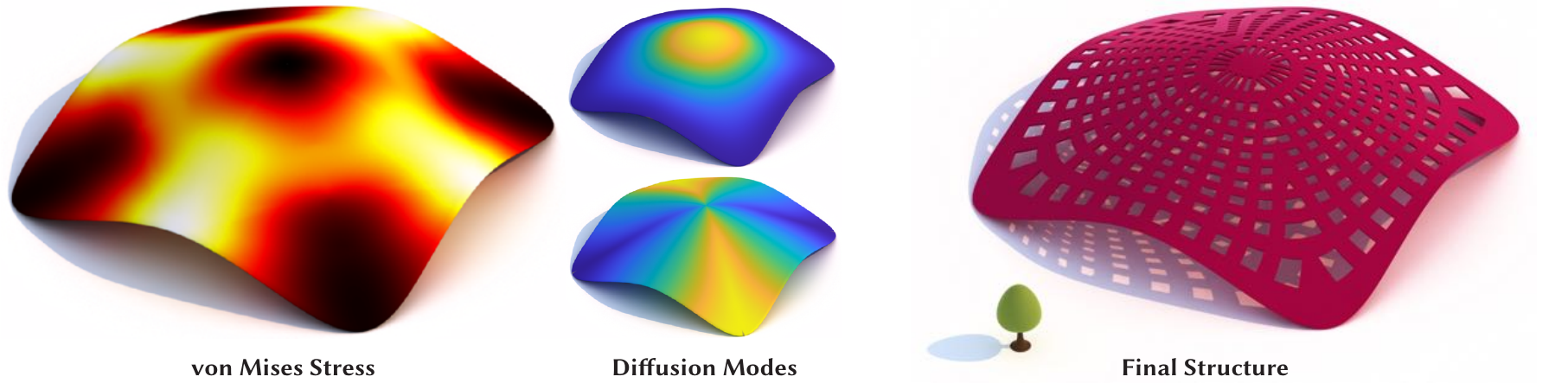}
  \centering
  \caption{Using internal stress directions (left) to drive anisotropic diffusion (middle)
  produces structurally influenced, aesthetically pleasing geometry (right, tree for scale).}
  \label{fig:teaser}
\end{teaserfigure}

\author{Abhishek Madan}
\affiliation{
  \institution{University of Toronto}
}
\email{amadan@cs.toronto.edu}
\author{Alec Jacobson}
\affiliation{
  \institution{University of Toronto}
}
\email{jacobson@cs.toronto.edu}
\author{David I.W. Levin}
\affiliation{
  \institution{University of Toronto}
}
\email{diwlevin@cs.toronto.edu}


\begin{abstract}
  We present \emph{Diffusion Structures}, a family of resilient shell structures from the eigenfunctions of a pair of novel diffusion operators. 
  This approach is based on Michell's theorem but avoids expensive non-linear optimization with computation that amounts to constructing and solving two generalized eigenvalue problems to generate two sets of stripe patterns. 
  This structure family can be generated quickly, and navigated in real-time using a small number of tuneable parameters.
\end{abstract}

%
%


%
%


\maketitle

\section{Introduction}

\begin{figure}
    \includegraphics[width=\columnwidth]{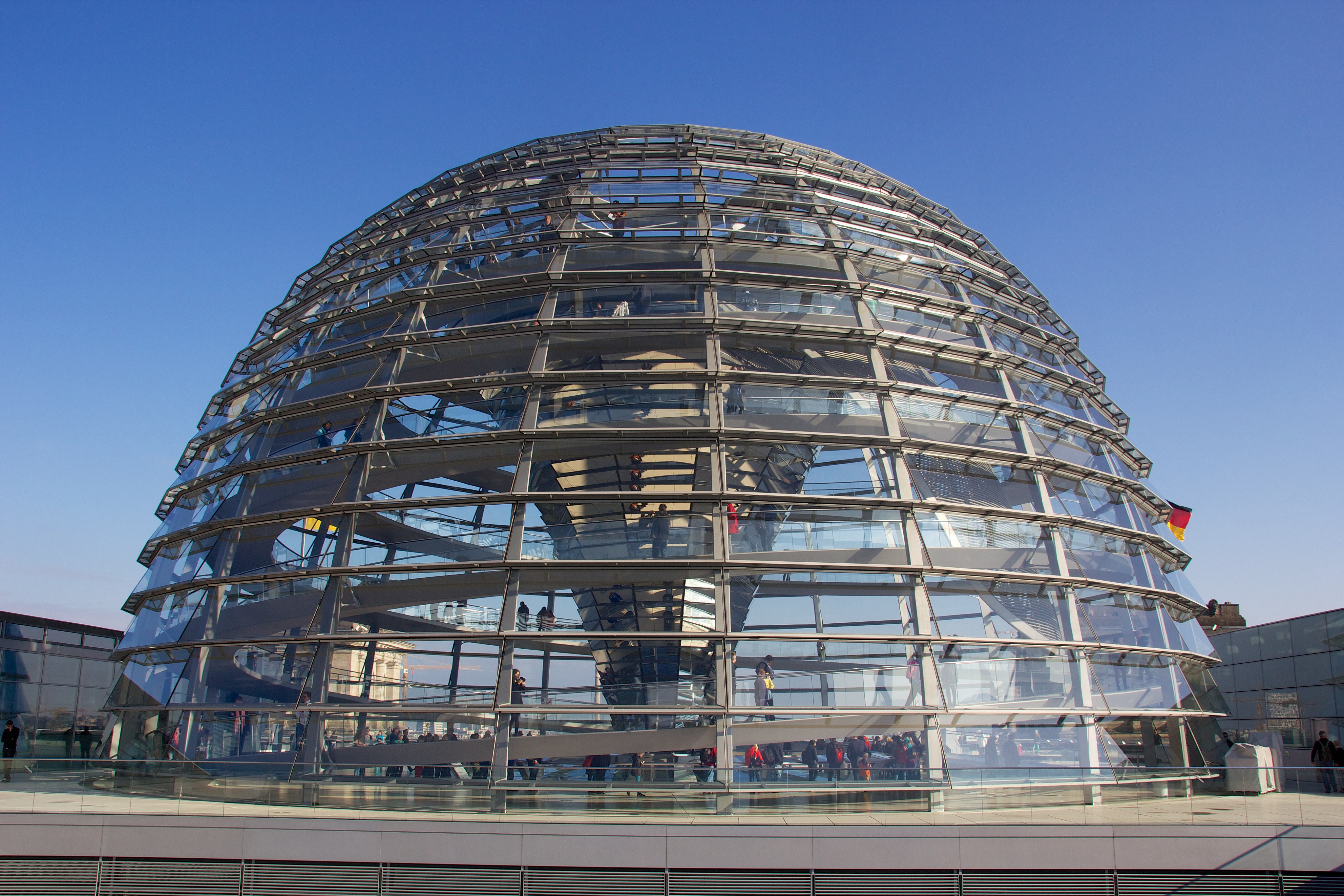}
    \caption{The Reichstag Dome in Berlin, Germany. The metal features are both aesthetic and performative --- they help resist radial and circumferential stresses.}
    \label{fig:reichstag}
\end{figure}

Structurally performative design attempts to strike a balance between the aesthetic and the structurally sound~\cite{tam2015stress}. 
The Reichstag dome~(Fig.~\ref{fig:reichstag}) provides a stunning example of this approach --- its metal and glass structure is both beautiful and resists radial and circumferential forces. 
Designs like this inspire our work on \emph{Diffusion Structures}, shape spaces tailored towards interactively designing performative geometries.

Diffusion Structures are topologically complex surface structures created from input surface geometry. 
Diffusion Structures proceed from the structural engineering premise that, for any structure, it is sufficient to analyze a single, dominant loading condition~\cite{sandaker2013structural}.
In the static load case, due to indeterminacy, there is a vast landscape of potential structures that can stably support the applied forces.
Our goal is to define a compact shape space which we can efficiently explore to find visually pleasing geometries.

Our approach differs from a standard structural optimization problem.
Structural optimization algorithms define optimality via an energy function (e.g minimum compliance) and then optimize over object shape, topology or both to minimize this objective. 
Variants of these approaches, which respect aesthetic guidance via constraints or cost function modifications, also exist. 
A consequence of this optimization-based providence is that output consists of only a single solution for a given loading scenario. 
This is satisfying from an engineering and mathematical perspective but less practical from a design exploration point-of-view.

Rather than treating structural optimality as a hard constraint and aesthetics as a secondary objective, we flip this
perspective and provide a constructive interface where users explore a shape space informed by the underlying physics via Michell's theorem~\cite{michell1904}. 
The intuitive interpretation of Michell's theorem is that material should be aligned with the principal stress directions of the domain.
Guided by this interpretation, we create structure by allowing material to diffuse through a stress field --- hence the term Diffusion Structure.
Crucially, creating Diffusion Structures requires the solution of only one linear system and two sparse eigensolves as a preprocess. 
Because efficient algorithms exist to perform these operations, a Diffusion Structure shape space can be initialized in a matter of seconds.
Once initialized, the Diffusion Structure topology can be updated in realtime using a simple fragment shader running on the graphics processing unit (GPU).
These features culminate to allow for efficient exploration of the structural space using only a handful of parameters. 

By prioritizing ease-of-exploration over exact structural optimality, Diffusion Structures provide an expressive prototyping
mechanism --- producing structurally informed geometry that will be further reinforced as part of the downstream engineering process. 
In the remainder of this manuscript we will outline the Diffusion Structures algorithm and demonstrate its efficacy and expressibility on a number of structural design problems.

\section{Related Work}
Diffusion Structures fall into the broad class of methods for generating structural geometry, that is geometry intended to be used in a load bearing scenario.
The most common algorithms in this category are optimization-based, modulating either the shape of an object, its topology, or both to produce a result.
Shape optimization has been used to to make objects stand~\cite{prevost2013}, spin~\cite{baecher2014} and float~\cite{wang2016}, but their goals differ significantly from ours, which is to produce lightweight structures.

Structural optimization schemes share our high-level goal, however their methodology is quite different.
These algorithms generate optimal structures by sparsifying a graph of structural elements. 
If this graph is composed of edges which represent bar elements, we arrive at the Ground Structure Method~\cite{freund2004}. 
Ground Structure Methods have been successfully applied to optimize the bar structure of architecture scale designs~\cite{jiang2017}. 
They are powerful, but suffer from limitations stemming from difficulties in choosing the initial input graph,
a large number of optimization variables and the fact that they produce a single output. 
Diffusion Structures are generated from an input surface (open or closed) and thus avoid the initialization problem.
More importantly for our use case, Diffusion Structures do not optimize for a single output, but instead parameterize an 
efficiently explorable design space. This property places them in a distinct category from Ground Structure approaches. 

Topology Optimization methods use a finite element discretization as their computational graph~\cite{bendsoe2013}, of which the most often used is a hexahedral mesh/voxel grid (e.g in Langlois et al~\shortcite{langlois2016}).
Topopology Optimization schemes parametrize material distributions per element~\cite{bendsoe1989} or by using level-sets~\cite{wang2003}, and sparsify material distribution in the  domain to produce a final output.
Topology Optimization exhibits the ability to produce high-performance, structurally optimal parts at large-scale~\cite{liu2018}.
As with the Ground Structure Method, Topology Optimization generates a single optimal design.
Exploring the design space can be difficult stemming from performance difficulties which are ameliorated by limiting results
to low resolution~\cite{nobeljorgensen2016} or to 2D~\cite{chen2018}. 
These performance issues extend to optimization-based approaches for surfaces~\cite{gilureta2019} as well.
Alternately, the Diffusion Structure approach --- rooted in an approximate, geometric notion of structural soundness --- produces an efficiently explorable structure space 
which contains intricate, high-resolution surface topology. 

Unlike optimization-based approaches, geometric structural optimization schemes attempt to explicitly align material with principal stress directions (in keeping with Michell's theorem~\cite{michell1904}).
The standard approach mimics that of quad-meshing (for shells)~\cite{jakob2015} or hex-meshing (for volumes)~\cite{nieser2011} algorithms, which take as input or compute a guiding, orthonormal frame field and then integrate this field.
Methods of this type can produce wire mesh-like structures but often require complex algorithms to handle singularities in the resulting fields~\cite{arora2019,wu2019design,sagemanfurnas2019}.
Diffusion Structures share the geometric motivation of these methods but avoid issues with singularities by never computing or integrating a gradient field. 
Instead, the stress tensor anisotropy is built into the diffusion stage which can seamlessly flow material through isotropic stress tensors, which are the root of the problem.

Diffusion primitives for content creation have a long and successful history in computer graphics. 
Diffusion curves~\cite{orzan2008} and surfaces~\cite{takayama2010} are used to create dense color fields from sparse input primitives --- curves and surfaces respectively. 
Diffusion Structures differ in that the diffusion process is used flow material across the surface of a mesh. 
In this sense our use of the diffusion operator is more closely related to diffusion distance computation~\cite{crane2017}.

The method by which we generate our final structures is inspired by the stripe generation algorithm of Kn\"oppel et al~\shortcite{knoppel2015}. 
However, our stated goal of creating structurally-influenced geometry leads to a unique diffusion-based method which we will outline below. 

Our work shares a superficial similarity with the work of Zehnder et al, which generates ornamental curve networks on surfaces~\shortcite{Zehnder2016}.
However, their approach focuses on adding additional, stabilizing structure to user-designed curve networks while
our method generates all geometry in a structurally informed way.








\section{Method}
Diffusion Structures take, as \textbf{input}, an open or closed 2D manifold, represented
as a stacked vector of vertex positions, $\V\in\Real^{3|\V|}$ and triangles, $\T\in\Real^{|\T|\times3}$. 
The \textbf{output} is a Diffusion Structure which consists of the input geometry and a \emph{topology texture} that denotes presence or absence of material for every point on the geometry. 
Fig.~\ref{fig:teaser} shows an input Pavilion geometry (left) and the final Diffusion Structure (right). 



\subsection{Structural Analysis}
\label{sec:structure}

\begin{figure}[hb]
  \includegraphics[width=\columnwidth]{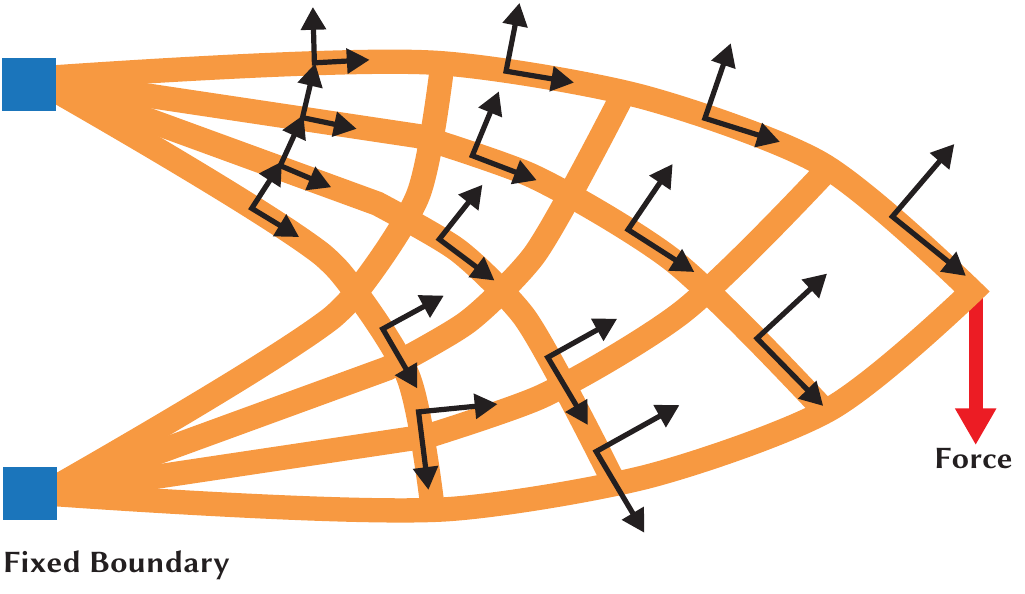}
  \caption{The principal stress directions for the classical Michell Truss and the truss structure itself.}
  \label{fig:tensors}
\end{figure}

Diffusion Structures inherent their notion of structural optimality from Michell~\shortcite{michell1904}. 
Michell showed that structures with optimal strength-to-weight ratios have material which follows the principal directions of the Cauchy stress tensors induced by loading conditions~(Fig.~\ref{fig:tensors}). 
While there is some debate about the true optimality of Michell Trusses~\cite{sigmund2016non}, the notion that structure should align with maximum load bearing directions 
is a useful guiding principal in architecture and structural engineering. 
Diffusion Structures are a computational approach to applying this principle on manifold surface geometry. 

Diffusion Structures can be built using any input stress tensor field. Our implementation computes this field using piecewise linear finite elements on $\V$ and $\T$.
Structural analysis is performed using a variational thin shell energy composed of the sum of bending and membrane energies.
For bending energy we use the discrete Wilmore Energy~\cite{wardetzky2007} and our membrane energy is the venerable St. Venant Kirchhoff (StVK) model. 

While our bending energy formulation is standard, we formulate our membrane energy extrinsically rather than the standard intrinsic approach~\cite{Bender2013}.
This has the advantage of allowing us to avoid the extra work of mapping a rectangular deformation gradient into the mesh tangent space.
Each triangle of our mesh is the medial surface of a thin slab of material and computing the per-slab membrane energy requires a per-triangle deformation gradient, $\F$.

\begin{figure}[hb!]
  \includegraphics[width=\columnwidth]{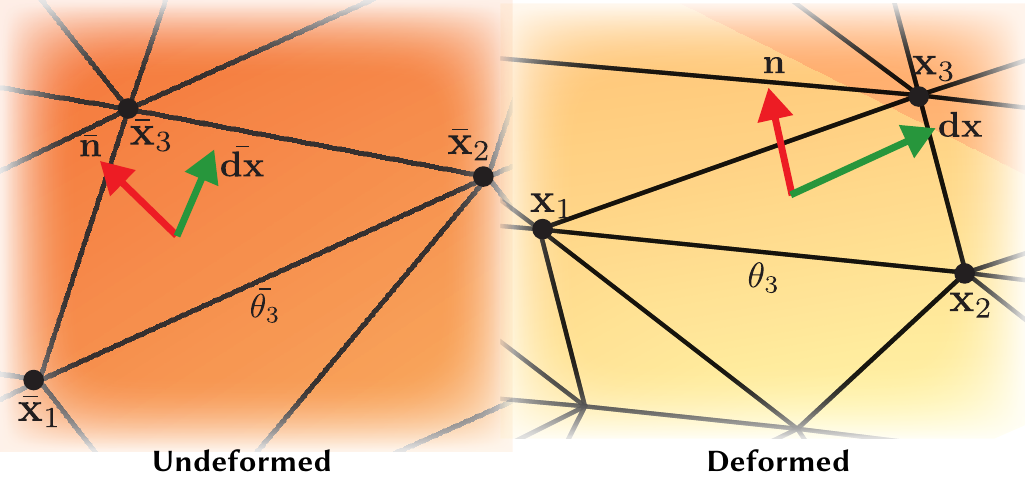}
  \caption{Our finite element setup for evaluating membrane deformation.}
  \label{fig:fem}
\end{figure}

Fig.~\ref{fig:fem} shows our finite element setup. $\F$ maps a vector in the undeformed, 3D space, $d\xun$, to its counterpart in the 3D, deformed space $d\xd$. 
For linear finite elements, the component of $d\xun$ in the tangent space deforms according to standard linear basis functions,
while the normal component remains unchanged (shells don't shear or stretch in the normal direction). 
These two assumptions give us a simple formula for per-triangle $\F\in\Real^{3\times3}$.

\begin{equation}
  \F = 
  \underbrace{\begin{pmatrix}
    \xd_1 & \xd_2 & \xd_3
  \end{pmatrix}
  \phi}_{\mbox{In Plane Stretch}}
  + \underbrace{\mathbf{n}\bar{\mathbf{n}}^T}_{\mbox{Normal}},
  \label{eq:F}
\end{equation} where $\xd_i \in \Real^3$ is the deformed position of the $i^{\text{th}}$ vertex,
$\mathbf{n}\in\Real^3$ is the unit length, deformed triangle normal,
and $\xun_i \in \Real^3$ and $\bar{\mathbf{n}}\in\Real^3$ are their undeformed counterparts.
The $3\times3$ matrix $\phi$ orthogonally projects $d\xun$ into the triangle tangent space using barycentric coordinates. 
Given the tangent vector matrix $Z=\begin{pmatrix} \xun_1 - \xun_3 & \xun_2 - \xun_3 \end{pmatrix}$, $\phi$ becomes

\begin{equation}
  \phi = 
  \begin{pmatrix}
    (Z^\top Z)^{-1}Z^\top \\
    -\mathds{1}^\top(Z^\top Z)^{-1}Z^\top)
  \end{pmatrix}, 
\end{equation}

where $\mathds{1}$ is the $1\times 3$ vector of ones. Intuitively, $\phi$ is performing a least 
squares projection of a point in 3D to the triangle tangent space by computing the 2D barycentric coordinates
for the nearest point on the triangle.

Using $F$, the total deformation energy for the triangle mesh becomes 

\begin{equation}
  \psi\left(\mathbf{\xd}\right) = \sum_{t=1}^{|\T|} \psi_{\scaleto{StVK}{3pt}}\left(\F_t\right)
    + \sum_{e=1}^{|E|} \psi_{\scaleto{Bend}{3pt}}\left(\theta_e\right), 
\end{equation} where $\mathbf{\xd}\in\Real^{3|\V|}$ is the stacked vector of per-vertex deformed positions, $|E|$ is the number of edges in the mesh and $\theta_e$ is the dihedral angle 
between triangles which share the $e^{th}$ edge. 

\begin{figure}[H]
  \includegraphics[width=\columnwidth]{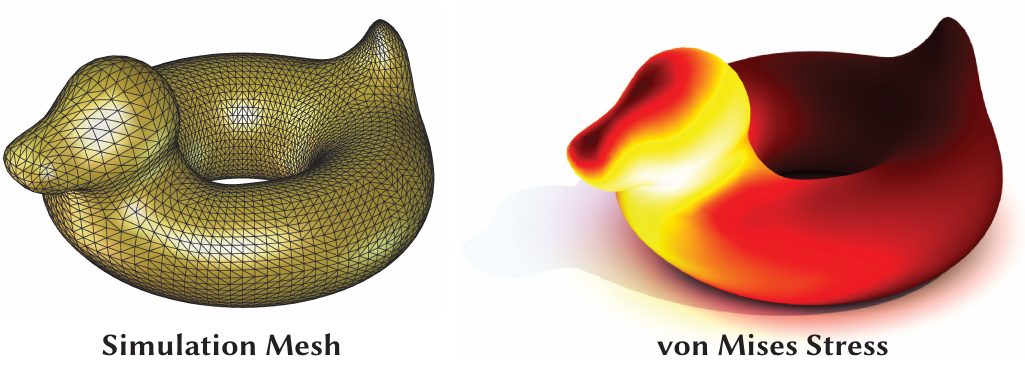}
  \caption{The simulation mesh for a Bob statuette (left) and the computed von Mises stress using our linear Finite Element solver (right).}
  \label{fig:sim_bob}
\end{figure}

Static analysis on structures undergoing infinitesimal deformation requires only that we minimize the quadratic approximation of 
this energy around the reference configuration, subject to Dirichlet boundary conditions. 
This approximation leads to a single linear system to be solved, which is given by

\begin{equation}
  H\mathbf{u} = \mathbf{f_{ext}},
\end{equation} where $\mathbf{u}\in\Real^{3|\V|}$  are per vertex displacements, $H = \frac{\partial^2 \psi}{\partial \xd^2}$ and $\mathbf{f_{ext}}$ are external forces.
In practice we apply Dirichlet boundary conditions by projecting fixed degrees-of-freedom out of the linear system.
Finally, we compute per-triangle Cauchy stresses from $\mathbf{u}$. Fig.~\ref{fig:sim_bob} shows the result of this stage of the algorithm.

\subsection{Anisotropic Diffusion}
\label{sec:diffusion}
Previous approaches to Michell truss-like structure generation follow an integration-based approach~\cite{arora2019} wherein
a pair of mutually orthogonal, stress aligned gradient fields are computed and then integrated to produce a map which aligns coordinate lines of a parametric domain
with these fields. Integration is performed by minimizing the well-known Dirichlet Energy. 

In the continuous case, minimizers of the Dirichlet Energy satisfy the Euler-Lagrange equations, $\nabla\cdot\nabla \alpha  = \nabla\cdot\mathbf{g}$.
Here $\alpha$ is the integrated scalar field and $\mathbf{g}$ is the gradient to be integrated. 
The major difficulty in these approaches lies in computing appropriate values of $\mathbf{g}$ in the presence of singularities, regions where the gradient can take on multiple values.
In structural analysis problems, these regions correspond to finite elements which contain isotropic stress tensors, and are surprisingly common.  

Alternatively, we can interpret the Euler-Lagrange equation as computing a distribution of material $\alpha$ that has been allowed to diffuse until it reaches static equilibrium.
Motivated by this interpretation, we replace Poisson integration with an anisotropic diffusion step that can more robustly handle isotropic stress tensors. 
Moving gradient information from the right-hand side of the Poisson equation, into the operator itself on the left-hand side, leads to the homogenous equation

\begin{equation}
  \nabla\cdot ( D \nabla \alpha ) = 0,
  \label{eq:anisodif}
\end{equation} where $D\in\Real^{3\times3}$ is the (extrinsic) diffusion tensor field which biases the directions in which material flows at a point in space. 
Our approach uses the Cauchy stress, $\sigma$ to create an approriate $D$ that aligns the flow of $\alpha$ with the principal stresses. 
We then use diffused fields to compute Diffusion Structures. 

\subsection{Diffusion Tensors from Stress}
\label{sec:metric}
An advantage of our simulation approach is that it produces $3\times3$ Cauchy stress tensors. Unfortunately, 
these tensors are not directly digestible by the diffusion equation due to their indefinite nature. 
This results from the fact that stress tensors not only measure the magnitude of force acting in a given direction but also whether that force
is a compressive or a tensile force. This compression/tension determination manifests as signed eigenvalues of the tensor.
However, we only care about the magnitude of force in each direction, not the compression/tension labelling, therefore it is 
sufficient to work with the absolute value of the principal stresses. Concretely we want to create a tensor, $D$,
that shares the eigenvectors of the stress tensor and whose eigenvalues are ordered by absolute value. 

To do this we examine the eigendecomposition of the Cauchy stress, $\sigma_i = \Sigma \Lambda \Sigma^\top$ and use it to create a new tensor $\sigma_i^\prime$ by 
rescaling the eigenvalue matrix $\Lambda$ into a new diagonal matrix $\Lambda^\prime$. Preserving the eigenvectors gives us 
$\sigma_i^\prime = \Sigma \Lambda^\prime \Sigma^\top$ which means that our diffusion directions will match the original principal stress directions.
To ensure that $\sigma_i^\prime$ meets the criteria above, it is sufficient to compute $\Lambda^\prime = |\Lambda|$.

\begin{figure}[t]
  \includegraphics[width=\columnwidth]{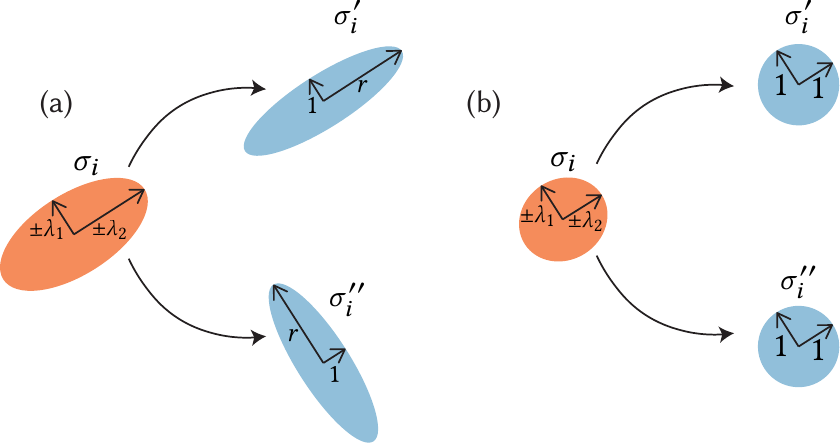}
  \caption{(a) If $\sigma_i$ is anisotropic, we map tangent eigenvalues $\pm \lambda_1$ and $\pm \lambda_2$ ($0 \le \lambda_1 \le \lambda_2$)
           to 1 and $r$ in $\sigma_i^\prime$ and vice versa in $\sigma_i^{\prime\prime}$.
           (b) On numerically isotropic tensors, we map both eigenvalues to 1 in both $\sigma_i^\prime$ and $\sigma_i^{\prime\prime}$.}
  \label{fig:rescale}
\end{figure}

Experimentally, we found that stress tensors with high degrees of anisotropy cause numerical issues in the resulting diffusion matrix.
We alleviate this by classifying tensors as either anisotropic or isotropic (based on the ratio of their two tangent plane eigenvalues, see Fig.\ref{fig:rescale}). 
We set the eigenvalues of isotropic tensors to be $1$ and prescribe a fixed anisotropy ration $r$ for anisotropic tensors.
Low values of $r$ have a smoothing effect, which converges to the isotropic case when $r = 1$, while high values of $r$ follow the data more closely.
We strike a balance by setting $r$ between $10$ and $100$, which worked well in all our experiments~(Table~\ref{tab:params}). Fig.~\ref{fig:anisotropy} shows the effect of $r$ on modes of the diffusion operator.
Although this rescaling discards eigenvalue magnitudes from the tensor data, this is acceptable because our objective is to follow the principal stress directions.



\subsection{Discretization}
We discretize equation~\ref{eq:anisodif} via the method of Andreux et al~\shortcite{andreux2014}, using our modified, per-face $\sigma_i^\prime$ as the diffusion tensor.
This results in the anisotropic diffusion operator $L_U$.
Counter-intuitively, material diffused using $L_U$ will move fastest in the direction orthogonal to the 
primary eigenvector of $\sigma^\prime_i$. This is because the anisotropy weight $r$ \emph{penalizes} gradients
in the primary eigenvector direction.
However, in \S\ref{sec:texture} we will show that such a flow generates structure
that is aligned with the primary eigenvectors.

Structure in one eigenvector direction is not enough to generate a connected object; we also require 
supporting structure running (locally) in the orthogonal direction. We synthesize this structure using 
an additional anisotropic diffusion operator $L_W$. This operator is constructed using diffusion tensors,
$\sigma^{\prime\prime}$, which penalize diffusion along the secondary eigenvector (Fig.~\ref{fig:rescale}). 

Without modification, solving the discrete analog of Equation~\ref{eq:anisodif} admits only the zero and constant solutions. We alleviate this by searching
for solution sets that are mutually orthonormal and computing a subset of low energy diffusion modes.
Given our two diffusion operators, we compute these modes by solving two sparse generalized eigenproblems
\begin{align}
  L_U U & = \lambda MU \\
  L_W W & = \lambda MW \\
  U^\top U &= I \\
  W^\top W &= I,
  \label{eq:eigv}
\end{align} where $M$ is the Voronoi mass matrix.

\begin{figure}[H]
  \includegraphics[width=\columnwidth]{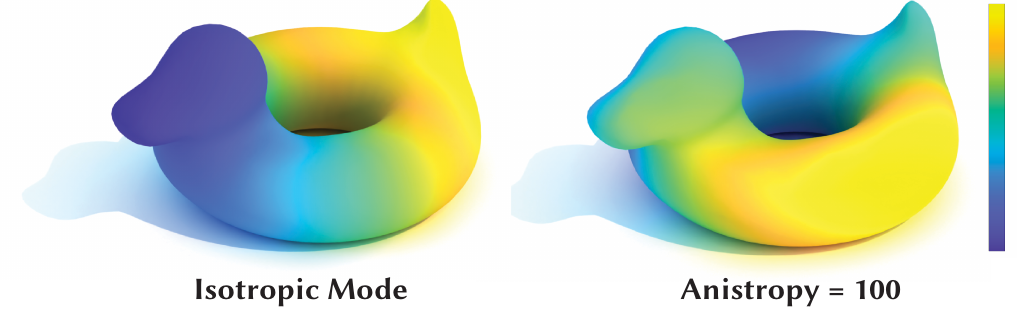}
  \caption{The effect of the anisotropy parameter. Left: the first non-constant mode of the isotropic diffusion operator. Right: the first non-constant mode of the anisotropic diffusion operator with r=100. High anisotropy encourages material to flow more readily in tensor aligned directions.}
  \label{fig:anisotropy}
\end{figure}

In the context of dynamic anisotropic diffusion, the modes $U$ and $W$ correspond to initial conditions closest to being in equilibrium
due to their already good distribution with respect to the underlying anisotropy~(Fig.~\ref{fig:anisotropy}).
This makes them excellent candidates from which to build Diffusion Structures. 
Using diffusion modes confers other advantages to our method: for instance, there is no need to select discrete points to diffuse
from, and we avoid issues with the exponential decay of solutions away from these initial points. 
Experimentally we observe that a good structure can be constructed from one of the first four computed modes,
making this calculation extremely efficient when using tools such as ARPACK. 

\subsection{Topology Texture Generation}
\label{sec:texture}

We construct our topology texture by generating structurally influenced stripe patterns from $U$ and $W$.
The user-selected modes of $U$ and $W$ are denoted $U_a$ and $W_b$, respectively, and $a$ and $b$ denote the mode's relative order in $U$ and $W$.
For ease of exposition, we assume for this section that $U$ and $W$ contain only a single diffusion mode.
For each triangle, $t$, we can construct a linear function  from either $U$ (resp. $W$) via barycentric interpolation.
Taking our lead from Kn\"oppel et al~\shortcite{knoppel2015}, we reinterpret this function, $\upsilon\left(\mathit{x}\right)$ (resp $\omega\left(\mathit{x}\right)$), as an angle 
and create material stripe patterns by evaluating a periodic function at each point on the surface:

\begin{align}
  s_1(\mathbf{x}) = & p(\alpha_{\scaleto{U}{3pt}} \upsilon(\mathbf{x})+\beta_{\scaleto{U}{3pt}}) - m_{\scaleto{U}{3pt}}, \\
  s_2(\mathbf{x}) = & p(\alpha_{\scaleto{W}{3pt}} \omega(\mathbf{x})+\beta_{\scaleto{W}{3pt}}) - m_{\scaleto{W}{3pt}}.
\end{align}

Here $p$ is a periodic function (we use a triangle wave with a period and amplitude of 1), $\alpha_{\scaleto{U}{3pt}}$ ($\alpha_{\scaleto{W}{3pt}}$) is the frequency,
$\beta_{\scaleto{U}{3pt}}$ ($\beta_{\scaleto{W}{3pt}}$) is the phase and $m_{\scaleto{U}{3pt}}$ ($m_{\scaleto{W}{3pt}}$) is the amplitude threshold.
Under this interpretation $\frac{\partial \upsilon}{\partial \mathit{x}}$ becomes the spatial angular velocity vector --- 
it denotes the direction in which $s_1$ changes the fastest.
Recall that our $U$ modes change most quickly along directions aligned with the secondary eigenvectors of our diffusion tensors.
Conversely, stripe values are relatively more constant orthogonal to this direction. 
This leads to stripes that propogate along the primary eigenvectors of our diffusion tensors.
These directions, by construction, correspond to the principal stress directions, hence $s_1$ produces
structure aligned with the major load-carrying directions of our input manifold. 
For this reason we refer to $U$ as the \emph{major} modes and $s_1$ as the \emph{major} stripe pattern
while $W$ and $s_2$ become the \emph{minor} modes and stripe pattern respectively.

What remains is to synthesize the entire Diffusion Structure, $s$, from $s_1$ and $s_2$, which we do via an implicit union,
computed as $s = \{ \mathbf{x} \mid \max(s_1(\mathbf{x}), s_2(\mathbf{x})) > 0 \}$.

\begin{figure}[h]
  \includegraphics[width=\columnwidth]{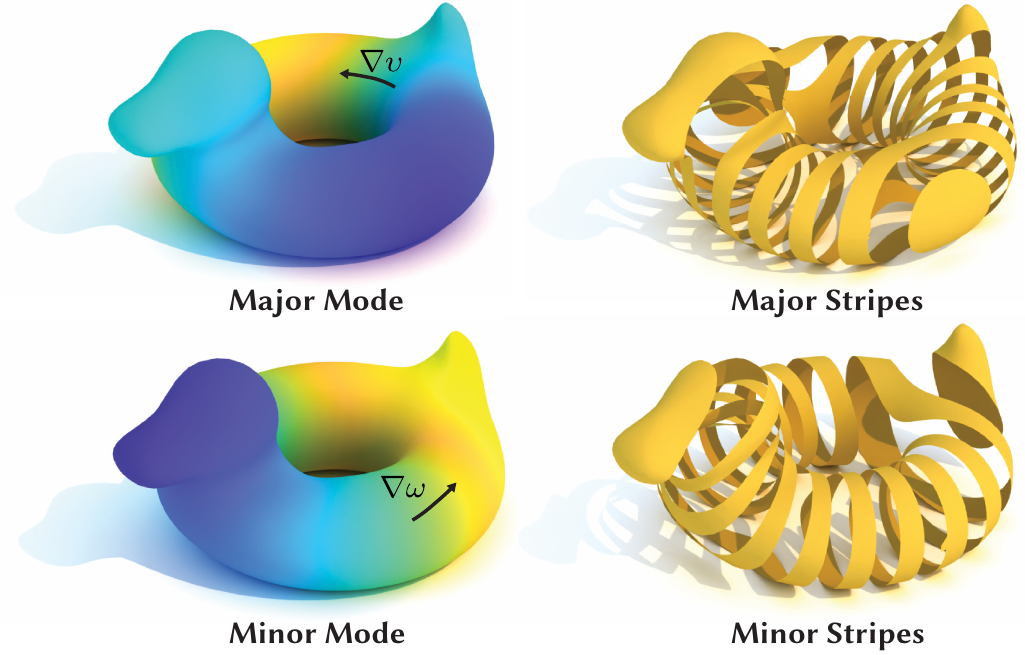}
  \caption{Anisotropic diffusion modes (left) and their corresponding stripe patterns (right) generated by our method. Each stripe runs perpendicular to the gradient of its respective mode (green arrows), while the pattern propagates along the gradient.}
  \label{fig:modes}
\end{figure}

Each tuple, $(\alpha_{\scaleto{U}{3pt}},\alpha_{\scaleto{W}{3pt}},\beta_{\scaleto{U}{3pt}},\beta_{\scaleto{W}{3pt}},m_{\scaleto{U}{3pt}},m_{\scaleto{W}{3pt}})$, defines a single Diffusion Structure in a 6-dimensional family of structurally influenced geometries. 
The remaining task, which we describe below, is to select a single Diffusion Structure from this structure space.
Fig.~\ref{fig:modes} (right) shows two sets of example stripe patterns generated by our method.

\subsection{Interactive Editing}
\label{sec:edit}

Since $\upsilon$ and $\omega$ are linear interpolations of per-vertex values $U$ and $W$, $s_1(\mathbf{x})$ and $s_2(\mathbf{x})$ can be very efficiently evaluated in a fragment shader.
Fragments outside $s$ are simply discarded.
This allows for interactive modification of the parameters $(\alpha_{\scaleto{U}{3pt}},\alpha_{\scaleto{W}{3pt}},\beta_{\scaleto{U}{3pt}},\beta_{\scaleto{W}{3pt}},m_{\scaleto{U}{3pt}},m_{\scaleto{W}{3pt}})$.

\begin{figure}[h]
  \includegraphics[width=\columnwidth]{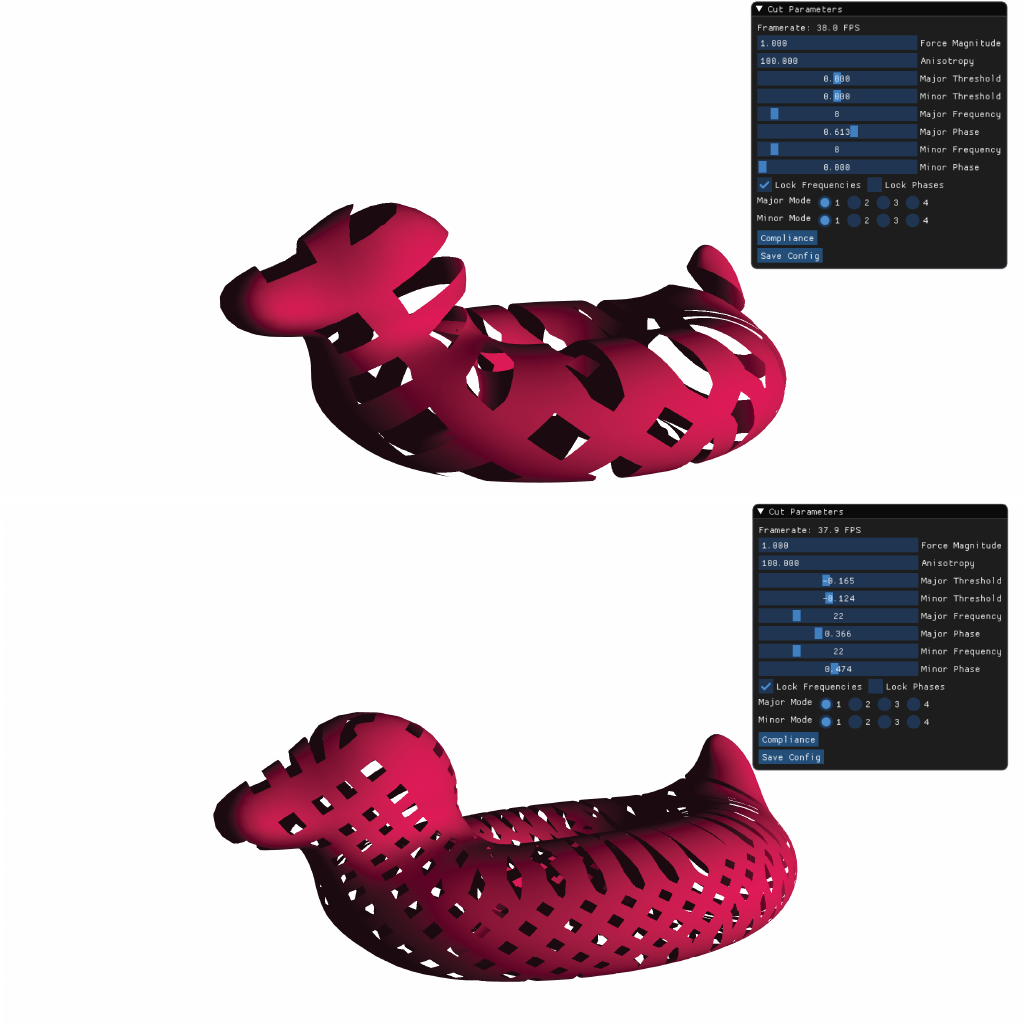}
  \caption{Modifying structure frequency and thickness in our interactive editor (unedited screenshots shown). In the case of the Bob mesh, these interactions happen at 37 frames per second.}
  \label{fig:editor}
\end{figure}

There are other useful parameters that can be tuned interactively, albeit not in real time (see Table~\ref{tab:timings}).
Since we solve an approximation of the statics problem, we also allow tuning the force magnitudes, which can be necessary if the original forces are too small to produce numerically anisotropic stress tensors.
Given the original $\mathbf{f_{ext}}$, we can create a new $\mathbf{f_{ext}}^\prime = \gamma\mathbf{f_{ext}}$ by a user-specified $\gamma$,
which corresponds to a new displacement vector $\mathbf{u}^\prime = \gamma\mathbf{u}$.
It is also useful to tune the anisotropy $r$ (see \S~\ref{sec:metric}), which only affects the diffusion matrices and not the statics problem.
Fig.~\ref{fig:editor} shows unedited screenshots from an interactive editing session.

\subsection{Triangle Mesh Generation}
\label{sec:mesh}
While an image-based approach to structure creation is invaluable for interactive editing, it is impractical for other uses.
Many downstream tools that might be used to post-process a Diffusion Structure expect meshes as input. 
We use adaptive upsampling of the input triangle mesh to produce a new triangulated output.
Specifically, we continuously upsample the input mesh triangles where isolines of $s_1(\mathbf{x}) = 0$ or $s_2(\mathbf{x}) = 0$
pass through more than once by adding new vertices on edge midpoints.
Next, we cut the mesh triangles along isolines, $s_1(\mathbf{x}) = 0$ and $s_1(\mathbf{x}) = 0$. 
Due to the upsampling step, no more than one isoline will pass through a triangle, which avoids aliasing artifacts.
Finally, we discard triangles outside of $s$. Due to the mesh slicing step, every triangle will be either entirely inside $s$ or entirely outside $s$. 
The final result is a triangle mesh that exactly matches the topology of the Diffusion Structure.

\section{Results}

We have used our interactive editor to generate a number of Diffusion Structures in both 2D and 3D.
All our results were generated using a 2015 MacBook Pro with a dual-core Intel i5 processor and 16 GB of RAM.
We used Eigen~\cite{eigenweb} for solving linear systems, and Spectra~\cite{spectraweb} for solving eigenvalue problems.
We will release our code to the public after publication.
Final parameters used for all results are shown in Table~\ref{tab:params} and timings, in seconds, are given in Table~\ref{tab:timings}.
\begin{table}[b]
    \centering
    \rowcolors{2}{white}{CornflowerBlue!25}
    \begin{tabular}{ l  r  r  r  r  r  r  r  r  r  r }
      \hline
      \textbf{Fig.} & $\mathbf{\gamma}$ & $\mathbf{r}$ & $\mathbf{a}$ & $\mathbf{b}$ & $\mathbf{m_{\scaleto{U}{3pt}}}$ & $\mathbf{m_{\scaleto{W}{3pt}}}$ & $\mathbf{\alpha_{\scaleto{U}{3pt}}}$ & $\mathbf{\alpha_{\scaleto{W}{3pt}}}$ & $\mathbf{\beta_{\scaleto{U}{3pt}}}$ & $\mathbf{\beta_{\scaleto{W}{3pt}}}$ \\
      \hline
      \ref{fig:2d}a & 5 & 100 & 1 & 1 & 0 & 0 & 8 & 8 & 0 & 0 \\
      \ref{fig:2d}b & 0.5 & 100 & 1 & 1 & 0 & 0 & 8 & 8 & 0 & 0 \\
      \ref{fig:2d}c & 1 & 100 & 1 & 1 & 0 & 0 & 8 & 8 & 0 & 0 \\
      \ref{fig:showcase}d & 10 & 100 & 1 & 2 & 0 & 0 & 5 & 5 & 0 & 0.15 \\
      \ref{fig:showcase}b & 1 & 100 & 1 & 1 & 0 & 0 & 10 & 10 & 0 & 0 \\
      \ref{fig:showcase}c & 10 & 100 & 1 & 2 & 0 & 0 & 15 & 15 & 0 & 0 \\ 
      \ref{fig:teaser} & 100 & 100 & 1 & 3 & 0 & 0 & 5 & 5 & 0 & 0 \\
      \ref{fig:camelposes}b & 10 & 100 & 1 & 1 & 0 & 0 & 8 & 8 & 0 & 0 \\ 
      \ref{fig:camelposes}c & 5 & 100 & 1 & 3 & 0 & 0 & 8 & 8 & 0 & 0.2 \\ 
      \ref{fig:camelposes}d & 10 & 100 & 1 & 3 & 0 & 0 & 8 & 8 & 0 & 0 \\ 
      \ref{fig:showcase}e & 1000 & 75 & 3 & 4 & 0 & 0 & 16 & 16 & 0 & 0 \\
      \ref{fig:showcase}g & 1 & 100 & 1 & 1 & 0 & 0 & 10 & 10 & 0 & 0 \\
      \ref{fig:showcase}h & 1 & 100 & 2 & 3 & 0 & 0 & 24 & 31 & 0 & 0 \\
      \ref{fig:showcase}f & 1000 & 10 & 3 & 3 & -0.16 & -0.13 & 4 & 7 & 1 & 0.57 \\
      \ref{fig:bridge} & 50 & 10 & 3 & 3 & 0 & 0 & 20 & 20 & 0 & 0 \\
      \ref{fig:showcase}a & 100 & 100 & 2 & 2 & 0 & 0 & 20 & 20 & 0 & 0 \\
      \hline
    \end{tabular}
    \caption{The parameters used to produce every result shown in the paper. Left to right: force magnitude, anisotropy, major (minor) mode, major (minor) threshold, major (minor) frequency, major (minor) phase.}
    \label{tab:params}
  \end{table}
What is immediately obvious is that Diffusion Structures are not just efficient, but fast to compute. 
Precomputation takes a matter of seconds on our almost five year-old testbed laptop, and we consistently record 
framerates of 30 frames-per-second or greater during interactive editing. This is without expending 
much effort on optimization. For instance, the performance of Eigen's linear solvers falls short of more
battle hardened alternatives such as Pardiso~\cite{pardisoweb}.

\begin{table*}[t]
    \centering
    \rowcolors{2}{white}{CornflowerBlue!25}
    \begin{tabular}{l r r r r r r r r}
      \hline
      \textbf{Fig.} & $\mathbf{|V|}$ & $\mathbf{|T|}$ & \textbf{Membrane(s)} & \textbf{Bending(s)} & \textbf{Statics(s)} & \textbf{Stress(s)} & \textbf{Diffusion(s)} & \textbf{Modes(s)} \\
      \hline
      \ref{fig:2d} & 7891 & 15855 & 0.076 & N/A & 0.085 & 0.029 & 0.028 & 0.231 \\
      \ref{fig:showcase}d & 2954 & 5904 & 0.0465 & 0.065 & 0.263 & 0.001 & 0.0132 & 0.075 \\
      \ref{fig:showcase}b & 3934 & 7868 & 0.083 & 0.117 & 0.628 & 0.001 & 0.020 & 0.143 \\
      \ref{fig:showcase}c & 5344 & 10688 & 0.105 & 0.198 & 1.887 & 0.002 & 0.027 & 0.654 \\ 
      \ref{fig:teaser} & 6273 & 12288 & 0.101 & 0.197 & 0.858 & 0.002 & 0.028 & 0.176 \\
      \ref{fig:camelposes} & 11717 & 23323 & 0.175 & 0.748 & 3.419 & 0.003 & 0.063 & 0.477 \\
      \ref{fig:showcase}e & 13387 & 26199 & 0.241 & 0.961 & 2.144 & 0.003 & 0.0732 & 0.502 \\
      \ref{fig:showcase}g & 15361 & 30718 & 0.253 & N/A & 1.342 & 0.004 & 0.081 & 0.647 \\
      \ref{fig:showcase}h & 20099 & 40194 & 0.282 & 1.982 & 3.951 & 0.004 & 0.107 & 0.709 \\
      \ref{fig:showcase}f & 23695 & 47398 & 0.418 & 2.920 & 17.722 & 0.007 & 0.129 & 1.417 \\
      \ref{fig:bridge} & 26114 & 52224 & 0.424 & 3.389 & 8.849 & 0.007 & 0.129 & 1.340 \\
      \ref{fig:showcase}a & 29304 & 58604 & 0.429 & 4.124 & 8.855 & 0.008 & 0.147 & 1.305 \\
      \hline
    \end{tabular}
    \caption{A summary of timings for every result shown in the paper. Only one set of timings is reported per unique mesh.}
    \label{tab:timings}
  \end{table*}

Even without these improvements, the interactive performance of Diffusion Structures is of significant novelty
when compared to structural optimization approaches. 
These methods have runtimes measured in minutes~\cite{wu2015system,gilureta2019}, hours~\cite{liu2018} or even days~\cite{langlois2016}.
While we stress that Diffusion Structures is a complement rather than a replacement to these methods, this performance comparison
shows that Diffusion Structures provide an, until now, missing interactive tool for exploring complex topological geometries in the context of 
structurally performative design. 

\begin{figure*}[htp]
    \includegraphics[width=\textwidth]{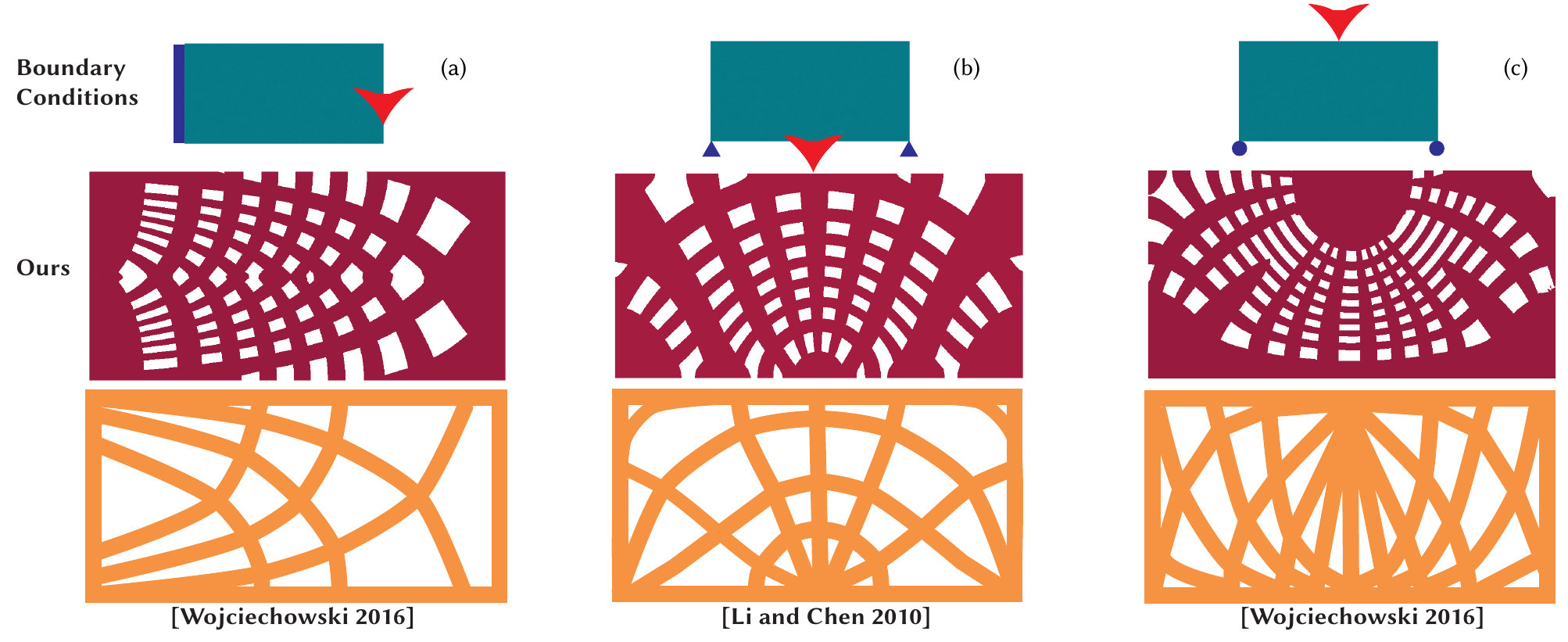}
    \caption{A variety of 2D examples. Neumann and Dirichlet boundary conditions shown in the top-left.
             (a) A hanging bar, held from the left and pulled down on the right.
             (b) Held at the corners and pulled from the bottom.
             (c) Pushed down and corners are allowed to ``roll'' horizontally but are fixed vertically.
           }
    \label{fig:2d}
\end{figure*}

Even when biased, diffusion is still an omnidirectional process and so Diffusion Structures are not guaranteed to align, exactly, with the underlying stress tensor field. 
To examine the magnitude of this effect, we first produced several 2D Diffusion Structures, based on standard loading cases from  Structural Engineering (\autoref{fig:2d}).
In all three cases, our results clearly follow the stress lines induced by the applied loads~\cite{Wojciechowski2016DeterminingOG,li2010beam}.
This demonstrates that Diffusion Structures are, in fact, structurally influenced. 
Interestingly, because Diffusion Structures diffuse material rather than explicitly generating curve networks,
they create additional joint-like structures near stress singularities. 
This is due to the isotropic nature of the stress field near these points and is similar to the type of 
feature more exact optimization schemes produce. 

\begin{figure*}[htp]
    \includegraphics[width=\textwidth]{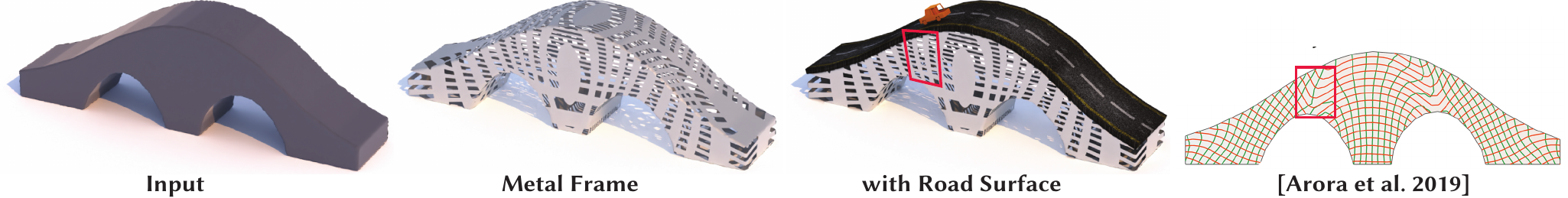}
    \caption{Diffusion Structures are used to create the metal frame for a bridge. Compared to previous work, the Diffusion Structure exhibits more regular behavior near divergent regions of the stress field.}
    \label{fig:bridge}
\end{figure*}

\autoref{fig:bridge} highlights another advantage of Diffusion Structures --- that they exhibit more regular behavior in divergent regions of the stress field than previous methods, which explicitly choose gradient directions for each element.
In regions which include divergent fields or singularities, there may not be an optimal choice, leading to artifacts.
Diffusion Structures avoid this issue by never making the choice, instead allowing material to diffuse isotropically at these points.
This leads to more visually pleasing output.

\begin{figure}[h]
    \includegraphics[width=\columnwidth]{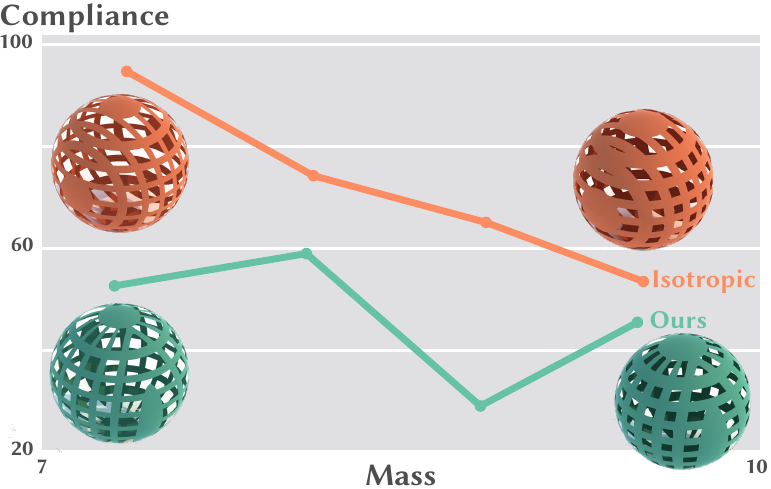}
    \caption{The anisotropy in Diffusion Structures provide them with lower compliances compared to isotropic structures with similar mass.}
    \label{fig:compliance3d}
\end{figure}

Next we evaluate the structural rigidity of Diffusion Structures in 3D using a unit sphere model. 
We compare the compliance of a Diffusion Structure to that of non-structurally influenced design for a variety of material volumes.
Note that our Diffusion Structure exhibits a lower compliance than the na\"{i}ve design, indicating that Diffusion Structures
generate geometry that is more resistant to applied loads~(\autoref{fig:compliance3d}). 

\begin{figure}[h]
    \includegraphics[width=\columnwidth]{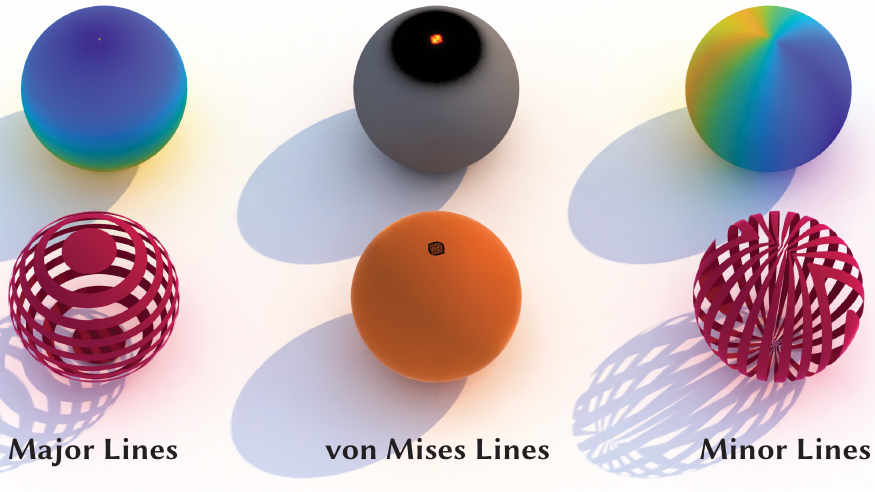}
    \caption{The von Mises stress alone (middle) only creates a single large stripe away from the poles. On the other hand, the diffusion modes (left, right) produce more regular stripe patterns across the surface.}
    \label{fig:vonmises}
\end{figure}

One might think of generating stripe patterns from the scalar von Mises stress field as an alternative to Diffusion Structures.
This approach produces unacceptable results since it is difficult to decompose the von Mises stress into two material families.
\autoref{fig:vonmises} shows the von Mises stripe pattern computed on a sphere fixed at the bottom pole and compressed at the top pole.
The two Diffusion Structure material families contain structure to resist radial and circumferential loads, while the single von Mises stripe pattern creates a single radial stripe due to its decay far from the poles of the sphere.
 
\begin{figure*}[htp]
    \includegraphics[width=\textwidth]{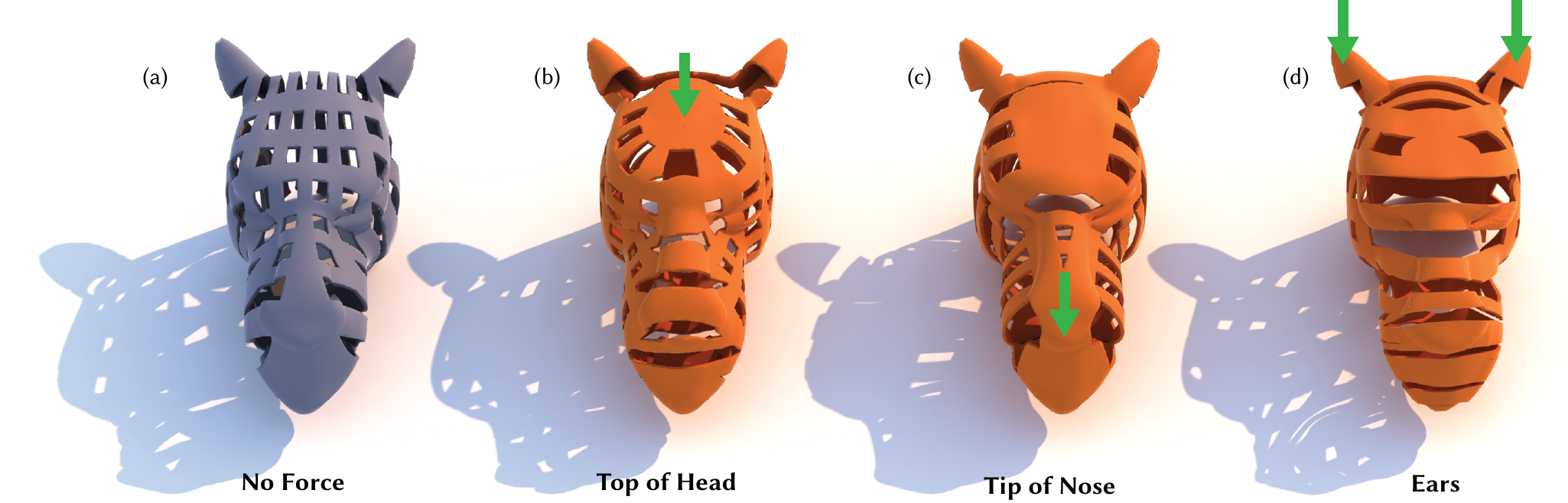}
    \caption{The stress tensors induce a variety of different patterns. No force creates a grid-like pattern, while pressing on the head, nose, and ears create radial patterns around those regions.}
    \label{fig:camelposes}
  \end{figure*}

Next we demonstrate how Diffusion Structures adapt to applied load on more complex meshes. 
The camel mesh in \autoref{fig:camelposes} features complex geometry and an open boundary at the base of the neck, which is fixed.
We apply several load cases to the mesh and observe the appearance of appropriate structural features for each. 
A radial web supports a load applied at the top of the head, a beam supports a load applied on the nose and
horizontal beams support a load applied across the head by pushing the ears down. 
We remind the reader that this is just one set of geometries, drawn from the space of Diffusion Structures, using our interactive editor.

\begin{figure*}[htp]
  \includegraphics[width=\textwidth]{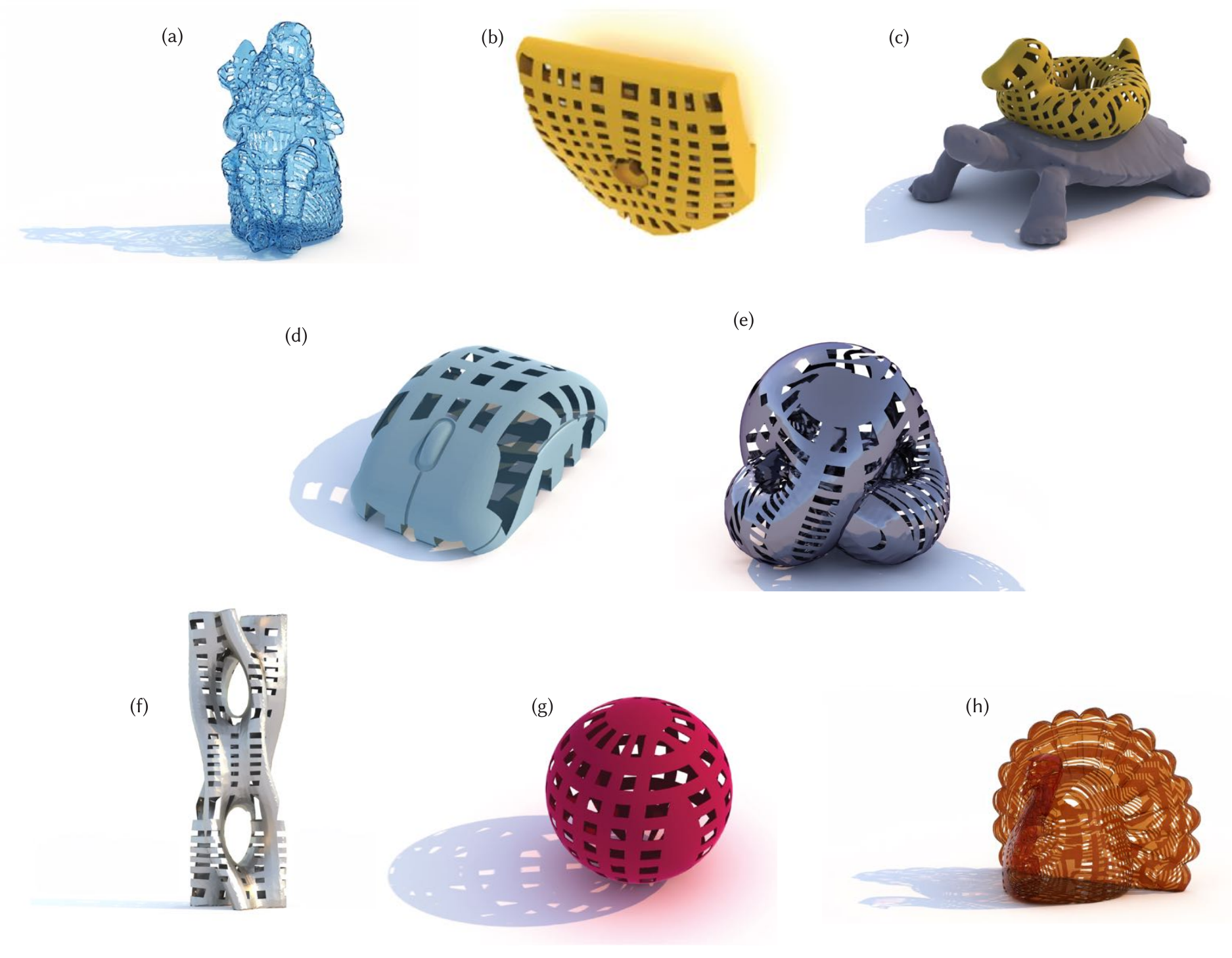}
  \caption{A showcase of Diffusion Structures.
           (a) A glass pirate, held at the feet and pushed on the back of the head.
           (b) A rock climbing hold, bolted into the wall and pressed down from the top.
           (c) Bob on top of the World Turtle, pressed down on the head.
           (d) A mouse being left-clicked.
           (e) Boy's Surface, pressed at the peak.
           (f) A twisty tower, pressed down from the top.
           (g) A simple sphere, compressed at one pole and held at the other.
           (h) A turkey, held at its base feathers and pushed on its head.
         }
  \label{fig:showcase}
\end{figure*}

Finally we show some additional examples of novel Diffusion Structures (\autoref{fig:showcase}).

\section{Discussion}
Diffusion Structures provide an efficiently explorable space of complex geometric structures and provide a compelling
companion to structural optimization techniques for generating performative geometry. 
We believe the relationship between these tools is symbiotic. 
While Diffusion Structures can quickly produce high-resolution, structurally influenced surface geometry,
the limitations of the approach mean it should not be considered a replacement to optimization techniques.

Our diffusion-based method for computing structure does not guarantee structural optimality,
nor does it take into account possible fracture. 
Diffusion structures also do not alter the thickness of the
resulting shape to improve load carrying behavior. 
This is why we refer to Diffusion Structures as structurally-influenced rather than structurally-optimal.
Similarly, Diffusion Structures do not consider the fabricability of the resulting geometry. 
All of these considerations demand post processing with additional tools further along the design and engineering
pipeline. Despite these limitations, we believe that the ease-of-exploration, coupled with the ability to generate intricate, structurally influenced geometries
makes Diffusion Structures an exciting addition to the collection of procedural geometry tools
used by architects and designers. 

We believe Diffusion Structures open doors to fascinating areas of future work. 
One obvious remaining challenge is to extend Diffusion Structures to the volumetric domain.
This requires addressing technical issues such as the generation of volumetric topology textures.
Exploring the Diffusion Structure space algorithmically is also an interesting avenue of future work and
would allow the fusion of Diffusion Structures with structural optimization. 
Since the space of Diffusion Structures is compact, this could lead to extremely efficient optimization schemes.
Last, but not least --- converting Diffusion Structures to fabricable blueprints is an important direction to explore. For similar geometries, this step is done manually or semi-manually, by large teams of engineers.
Automating this process will allow for more cost-effective deployment of Diffusion Structures into the real-world. 

\section{Conclusion}
We have presented Diffusion Structures --- topologically complex, structurally influenced shapes for architecture and design.
The strength of Diffusion Structures lies in their efficient computability which allows for
quick exploration of the structure space. 
This makes Diffusion Structures an exciting new tool in the arsenal of artists and designers 
working on performative design. 
We are excited to see what new designs will come about, powered by the ease-of-use and geometric complexity of
Diffusion Structures. 

\section{Acknowledgements}
We thank Rahul Arora, Sarah Kushner, John Kanji, and Josh Holinaty for their help in figure creation.
This work was funded in part by Discovery RGPIN-2017-05524, Accelerator RGPAS-2017- 507909, Connaught Fund 503114, and CFI-JELF: Canada Research Chairs.
The Pirate mesh is from TurboSquid
(https://www.turbosquid.com/3d-models/3d-model-of-decorative-pirate-zbrush/1084497),
and the Turkey mesh is from Thingiverse (https://www.thingiverse.com/thing:3957882).

\bibliographystyle{ACM-Reference-Format}
\bibliography{diffusion-structures}

\end{document}